# Crystal-Field–Driven Magnetoelectricity in the Triangular Quantum Magnet $CeMgAl_{11}O_{19}$

Sonu Kumar[1,2]*, Gaël Bastien[1], Maxim Savinov[3], Petr Proschek[1], Adam Eliáš[1], Karol Załęski[4], Małgorzata Śliwińska-Bartkowiak[2], Ross H. Colman[1], and Stanislav Kamba[3]†

*[1] Charles University, Faculty of Mathematics and Physics, Department of Condensed Matter Physics, Prague, Czech Republic*

*[2] Adam Mickiewicz University, Faculty of Physics and Astronomy, Department of Experimental Physics of Condensed Phase, Poznań, Poland*

*[3] Institute of Physics, Czech Academy of Sciences, Prague, Czech Republic*

*[4] Adam Mickiewicz University, NanoBioMedical Centre, Poznań, Poland*

*Corresponding author e-mail: sonu.kumar@matfyz.cuni.cz

†Corresponding author e-mail: kamba@fzu.cz

## Abstract

We report dielectric and magnetoelectric studies of single-crystalline $CeMgAl_{11}O_{19}$, a Kramers triangular magnet embedded in a polarizable hexaaluminate lattice. In zero magnetic field, the permittivity ε'(T) follows the Barrett law of a quantum paraelectric down to approximately 25 K, below which a broad minimum develops near 3 K without evidence of static ferroelectric or magnetic order. Application of magnetic fields up to 9 T shifts this minimum to higher temperatures and broadens it, evidencing a tunable magnetoelectric response. The magnetoelectric coupling was characterized using results from magnetization measurements. The anomaly temperature T*, extracted from the local minimum of ε'(T), exhibits a linear dependence on the squared magnetization $M^2$, consistent with the biquadratic magnetoelectric coupling allowed in centrosymmetric systems. This magnetoelectric effect, mediated by spin–orbit–entangled Kramers doublets interacting with a frustrated antipolar liquid, establishes $CeMgAl_{11}O_{19}$ as a prototype for exploring quantum magnetoelectricity in frustrated systems.

## I. Introduction

The possibility of interconverting electric and magnetic fields was already implicit in Pierre Curie's early symmetry arguments, but it was only in 1960 that the linear magnetoelectric (ME) effect was first observed experimentally by Astrov in $Cr_2O_3$ [1]. Since then, the concept of ME coupling has been greatly expanded to include higher-order effects beyond the linear regime. Symmetry analyses have shown that when the linear invariant $P_iM_j$ is forbidden by crystal symmetries—i.e., when time and space inversion are not simultaneously broken—the leading coupling terms become quadratic or biquadratic [2].

Microscopically, the magnetoelectric coupling can arise via three main mechanisms: symmetric exchange striction [3], inverse Dzialoshinskii-Moriya interaction [4, 5], and spin-dependent metal–ligand hybridization [6, 7]. The first mechanism originates from spin–lattice coupling in collinear spin systems and is generally the strongest. The other two, which are weaker, result from spin–orbit coupling in systems with non-parallel spins—typical of magnetically modulated phases. Linear ME coupling is not strictly limited to multiferroic systems; it has been observed, for example, in antiferromagnetic $Cr_2O_3$ [1], $DyFeO_3$ [8], and in field-induced states of the quantum magnet $TlCuCl_3$ [9].

Triangular-lattice antiferromagnets (TLAFs) represent a paradigmatic class of geometrically frustrated systems in which classical Néel order is destabilized by the inherent triangular connectivity, giving rise to a variety of exotic ground states and excitations. In spin-½ TLAFs, additional ingredients such as anisotropic exchange or further-neighbor couplings can destabilize magnetic order and promote candidate quantum spin liquids (QSLs). Proposed examples range from gapless algebraic Dirac liquids in organic salts and rare-earth chalcogenides [10, 11] to spinon Fermi-surface states in Yb-based systems [12, 13]. Numerical studies further reveal that exchange anisotropies (e.g., XXZ or Kitaev-like terms) can stabilize novel phases including chiral and dual QSLs [14].

A natural, albeit less-explored, analogue exists in the dielectric sector. In the hexaaluminates $AAl_{12}O_{19}$ (A = Ca, Sr, Eu, Ba, Pb) and $LnMAl_{11}O_{19}$ (Ln = La–Gd, M = Mg, Zn) with the magnetoplumbite structure, an Ising-like triangular lattice of electric dipoles forms, whose nearest-neighbor Coulomb interactions are antipolar [15–17]. First-principles calculations predicted that these dipoles should be frustrated in the same sense as the TLAF, with second-neighbor terms selecting stripe-type antiferroelectric order at very low temperature [18]. Experiments on Zn-doped $CaAl_{12}O_{19}$, $SrAl_{12}O_{19}$, and related phases confirmed sizable off-centering of $Al^{3+}$ ions and robust dipole moments [19].

The hexaaluminates $LnMgAl_{11}O_{19}$ (Ln = Ce–Gd) and $EuAl_{12}O_{19}$ harbor both a triangular lattice of rare-earth magnetic ions and a triangular lattice of electric dipoles. Recent characterization of the dielectric and magnetic properties of $EuAl_{12}O_{19}$ bridges these ideas by exhibiting both a frustrated antipolar "classical electric-dipole liquid" and quasi-two-dimensional ferromagnetism of $Eu^{2+}$ ($4f^7$, S = 7/2) [20]. The Eu moments order ferromagnetically below $T_C \approx 1.3$ K, whereas the dipole subsystem remains dynamically disordered down to at least 0.3 K and shows a broad, Arrhenius-type relaxation that freezes only as T → 0 K. A second-order transition at $T_S$ = 49 K marks the onset of slow correlated dipole dynamics, but X-ray and SHG measurements confirm that the global structure remains centrosymmetric.

Replacing $Eu^{2+}$ by $Ce^{3+}$ yields $CeMgAl_{11}O_{19}$, in which strong spin–orbit coupling selects an almost pure |±5/2⟩ Kramers doublet, giving rise to an effective $S_{eff}$ = ½ triangular lattice with pronounced XXZ anisotropy. Magnetization, specific-heat, and inelastic-neutron studies have shown that (i) no long-range magnetic order occurs down to 30 mK, (ii) the ratio $J_z/J_\perp$ lies near the ferromagnetic–120° quantum-critical point, and (iii) an excitation continuum is present [21, 22]. The same magnetoplumbite framework therefore hosts frustrated electric dipoles and a nearly critical quantum magnet on top of one another, opening a route to correlated magnetoelectric (ME) states in which the Landau-allowed $P^2M^2$ invariant couples the two sectors directly [2, 23].

Here, we present a comprehensive study of $CeMgAl_{11}O_{19}$, focusing on its dielectric and magnetoelectric properties under varying temperature and magnetic field. We demonstrate its quantum-paraelectric behavior and identify a biquadratic magnetoelectric coupling mediated by spin–orbit-entangled Kramers doublets. Supported by Landau theory and experimental evidence, these results highlight $CeMgAl_{11}O_{19}$ as a promising platform for exploring quantum magnetoelectricity in frustrated systems.

## II. Methods

Single crystals of $CeMgAl_{11}O_{19}$ were synthesized using a two-step process combining solid-state reaction with optical floating-zone (OFZ) crystal growth. High-purity binary oxides—$CeO_2$, MgO, and $Al_2O_3$ (all 99.99%)—were used as starting materials. These precursors were first calcined in air at 800 °C for 24 h to eliminate moisture and carbonate impurities. The calcined powders were weighed in stoichiometric proportions, mixed thoroughly, and ground to ensure uniformity.

The resulting powder was pressed into cylindrical rods (6 mm diameter, 100 mm length) under a quasi-hydrostatic pressure of 2 tons for 15 minutes. These rods were then sintered in air at 1200 °C for 72 h to promote complete solid-state reaction and improve densification. Crystal growth was performed in a four-mirror OFZ furnace under flowing air with a slight overpressure (1 atm) and an air flow rate of 3 L $min^{-1}$ to suppress evaporation. The feed and seed rods were counter-rotated at 30 rpm to stabilize the molten zone, and a pulling rate of 2 mm $h^{-1}$ was used. The resulting ingot was transparent white in color and contained multiple large grains with visible grain boundaries. These grains were isolated using a wire saw or mechanical cleavage and confirmed to be single crystals via backscattered Laue X-ray diffraction.

Further details on the synthesis and single-crystal structural refinement are provided in our previous publications [24–26].

For dielectric measurements, large single-crystal grains with minimal visible defects were selected. Each crystal was polished along the crystallographic c-axis to obtain parallel faces. Gold electrodes were deposited on the polished surfaces by thermal evaporation to ensure reliable electrical contacts.

The dielectric permittivity ($\varepsilon'$) was measured using an Alpha-AN high-performance impedance analyzer (Novocontrol Technologies) over the frequency range 1 Hz – 1 MHz. The sample, with a diameter of 7.2 mm and thickness of 0.51 mm, was mounted in a $^3$He cryostat (Cryogenic Ltd.) equipped with a 9 T superconducting magnet, with both electric and magnetic fields applied parallel to the c-axis. Measurements were performed while varying the temperature from 0.3 K to 300 K in zero field and from 0.3 K to 30 K under magnetic fields from 0 T to 9 T.

At each temperature, the sample was zero-field cooled. After reaching thermal stability, a magnetic field was applied and frequency-dependent dielectric spectra were recorded. This protocol ensured high reproducibility and controlled evaluation of the field- and temperature-dependent dielectric response.

The dc magnetic susceptibility was measured using a Quantum Design Magnetic Property Measurement System (MPMS) based on a superconducting quantum interference device (SQUID).

## III. Results

Figure 1(a) shows the temperature dependence of the permittivity $\varepsilon'(T)$ below 160 K. On cooling, $\varepsilon'(T)$ increases, reaching a maximum near 25 K, before decreasing slightly toward 3 K and increasing again toward the lowest measured temperature (0.35 K). Above 20 K, $\varepsilon'(T)$ follows the Barrett relation, which is typical of quantum paraelectrics or incipient ferroelectrics [27, 28]:

$$\varepsilon'(T) = \varepsilon\infty + C / [(T_1 / 2) \coth(T_1 / 2T) - T_0], \qquad (1)$$

with fitted parameters $\varepsilon\infty = 20.04 \pm 0.02$, $C = 263.3 \pm 2$ K, $T_1 = 160.6 \pm 0.6$ K, and $T_0 = 4.18 \pm 0.7$ K.

Here $\varepsilon\infty$ represents the background permittivity, $C$ is a Curie-like constant setting the strength of the soft polar mode, $T_1$ marks the temperature below which quantum fluctuations become important, and $T_0$ is the classical transition temperature. A positive $T_0$ = 4.18 K indicates quantum-paraelectric behavior. Fitting only the data below 160 K yields $T_0$ = −3.5 K, typical for incipient ferroelectrics. In either case, $|T_0| \approx 0$ K indicates strong low-temperature quantum fluctuations of electric dipoles originating from $AlO_5$ bipyramids, analogous to $EuAl_{12}O_{19}$ [20].

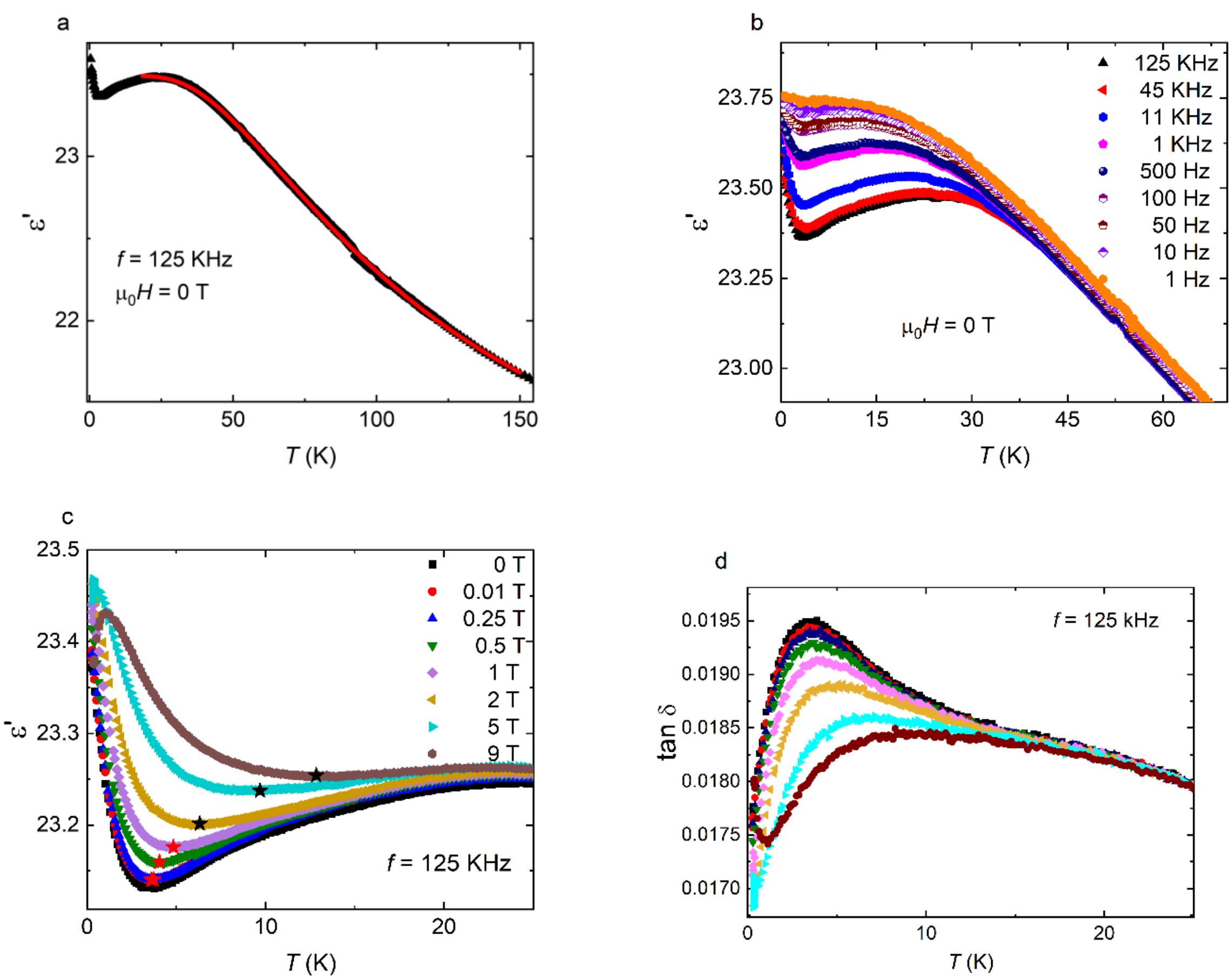

***Figure1.***

*(a) Temperature dependence of the permittivity ε'(T) at 125 kHz (black symbols) fitted with the Barrett formula [Eq. (1)] (solid red line). (b) ε'(T) measured at frequencies from 1 Hz to 125 kHz in zero magnetic field, showing clear dielectric dispersion below 30 K. (c) ε'(T) measured at 125 kHz under various magnetic fields. Red stars mark the anomaly temperatures T* extracted directly from the zero crossings of dε'/dT (for 0.1–1 T), while black stars indicate T* at higher fields (2–9 T) as predicted by the $P^2M^2$ Landau–Brillouin model. (d) Temperature and magnetic-field dependence of the dielectric loss measured at 125 kHz. The colors of the curves correspond to the magnetic fields shown in panel (c).*

As in other quantum paraelectrics, the temperature dependence of ε'(T) likely reflects weak softening of a polar phonon upon cooling. The low-temperature maximum in ε'(T) can arise from (i) coupling between the soft optical phonon and an acoustic branch, as recently confirmed in $SrTiO_3$ [29], or (ii) coupling between the soft mode and lattice defects [30, 31]. Because $CeMgAl_{11}O_{19}$ exhibits much lower permittivity than $SrTiO_3$, its lowest optical phonon must lie at relatively high frequency (≈ 3 THz in $EuAl_{12}O_{19}$ [20]), making mechanism (i) unlikely. The observed maximum is therefore most plausibly caused by mechanism (ii). This interpretation is supported by the frequency dependence of ε'(T) below 30 K [Fig. 1(b)], where the minimum near 5 K observed at 1 Hz diminishes at higher frequencies. Such dielectric dispersion can also reflect correlated dynamics of antipolar nanoclusters similar to those in $EuAl_{12}O_{19}$ [20].

When magnetic fields are applied, the broad dip in ε'(T) shifts steadily to higher temperatures and broadens further [Fig. 1(c)]. This field-tunable behavior persists up to 9 T, providing clear evidence of magnetoelectric (ME) coupling. In contrast, $EuAl_{12}O_{19}$ shows strong dipolar frustration but no measurable ME response [20].

Magnetization measurements [Fig. 2(a)] reveal pronounced easy-axis anisotropy consistent with previous work [21]. For H ‖ c, M(H) at 2 K rises sharply and saturates above ≈ 4 T, whereas for H ⊥ c no saturation occurs up to 7 T. Because magnetic interactions are weak ($J_\perp \approx 0.05$ meV [21]), the magnetization in the 2–10 K range is well described by a simple Brillouin-function form for an effective J = ½ moment:

$$M(H, T) = M_{sat} \tanh(g\, \mu_B\, H / 2\, k_B\, T), \qquad (2)$$

yielding $M_{sat} = 1.69\ \mu_B$ per Ce and g = 3.9, with uncertainties below 1%.

$CeMgAl_{11}O_{19}$ crystallizes in the centrosymmetric and time-reversal-invariant space group $P6_3/mmc$, with local $D_3h$ symmetry at the rare-earth site. Consequently, linear ME coupling ($\sim P_c\, M_c$) is symmetry-forbidden, while the lowest-order allowed term is biquadratic ($\sim P_c^2\, M_c^2$). In a Landau free-energy framework, this term does not induce a net polarization but renormalizes the polar stiffness:

$$\alpha_P^{eff}(T, H) = \alpha_P(T) + 2\, \gamma\, C\, M^2(H, T). \qquad (3)$$

Its experimental signature is a field-dependent shift of the dielectric anomaly temperature T*, with little change in overall permittivity magnitude or magnetic susceptibility. The characteristic fingerprint of biquadratic ME coupling is a linear scaling of T* with $M^2$. Such renormalizations have been identified in classical multiferroics (e.g., $DyFeO_3$ [8]), in spin-phonon-coupled quantum paraelectrics such as $EuTiO_3$ [2, 32], and in incipient ferroelectrics like $SrTiO_3$ and $KTaO_3$ [28].

To test this quadratic coupling, we analyzed ε'(T) at 125 kHz for fields 0.1–1 T, where the permittivity dip remains sharp. Each curve was smoothed with a 20-point Savitzky–Golay filter and differentiated between 2 and 7 K; the condition (dε'/dT) = 0 defines T*. The extracted anomaly temperatures are:

| H (T) | T* (K) |
|---|---|
| 0.00 | 3.6 |
| 0.10 | 3.6 |
| 0.25 | 3.7 |
| 0.50 | 4.0 |
| 1.00 | 4.8 |

For each {H, T*} pair, we computed the isothermal magnetization using Eq. (2). Plotting T* against $M^2$ [Fig. 2(b)] yields an almost perfect linear relation:

$$T^* = 3.6 + 6.2\ M^2\ (\mathrm{K}). \qquad (4)$$

This exact proportionality between T* and $M^2$ confirms experimentally that the dominant ME invariant in $CeMgAl_{11}O_{19}$ is of the $P^2M^2$ form, as required by its centrosymmetric, time-reversal-invariant crystal symmetry.

Combining the parameters from Eqs. (2) and (4), the relation $T^* = T_0 + a\ M^2$ and the Brillouin-function dependence M(H, T*) lead to an implicit expression:

$$T^*(H) = 3.6 + 17.6\ \tanh^2(1.3\ H\ /\ T^*(H)). \qquad (5)$$

Numerical solutions of Eq. (5) give T*(H) ≈ 6.3 K at 2 T, 9.7 K at 5 T, and 12.8 K at 9 T, consistent with Fig. 2(b). The T* values extracted directly from dε'/dT = 0 (0–1 T) and those predicted by the $P^2M^2$ relation (2–9 T) together accurately map the dielectric anomaly across the full field range [Fig. 1(c)].

**IV. Discussion**

$CeMgAl_{11}O_{19}$ exhibits the defining characteristics of a quantum paraelectric: its permittivity ε'(T) follows the Barrett law down to approximately 25 K [Fig. 1(a)], and no ferroelectric or magnetic ordering is observed down to 0.3 K. Below about 25 K, a distinct anomaly appears as a broad minimum in ε'(T) near 3 K [Fig. 1(a)], signifying that quantum fluctuations suppress a classical ferroelectric transition [33, 28]. When a magnetic field is applied, this minimum shifts steadily to higher temperatures and broadens [Fig. 1(c)], providing direct evidence of magnetoelectric (ME) coupling [34]. This behavior contrasts sharply with that of $EuAl_{12}O_{19}$, a structural analogue in which frustrated $AlO_5$ dipoles contribute to the dielectric constant but remain decoupled from $Eu^{2+}$ spins, producing no field-dependent dielectric response [20].

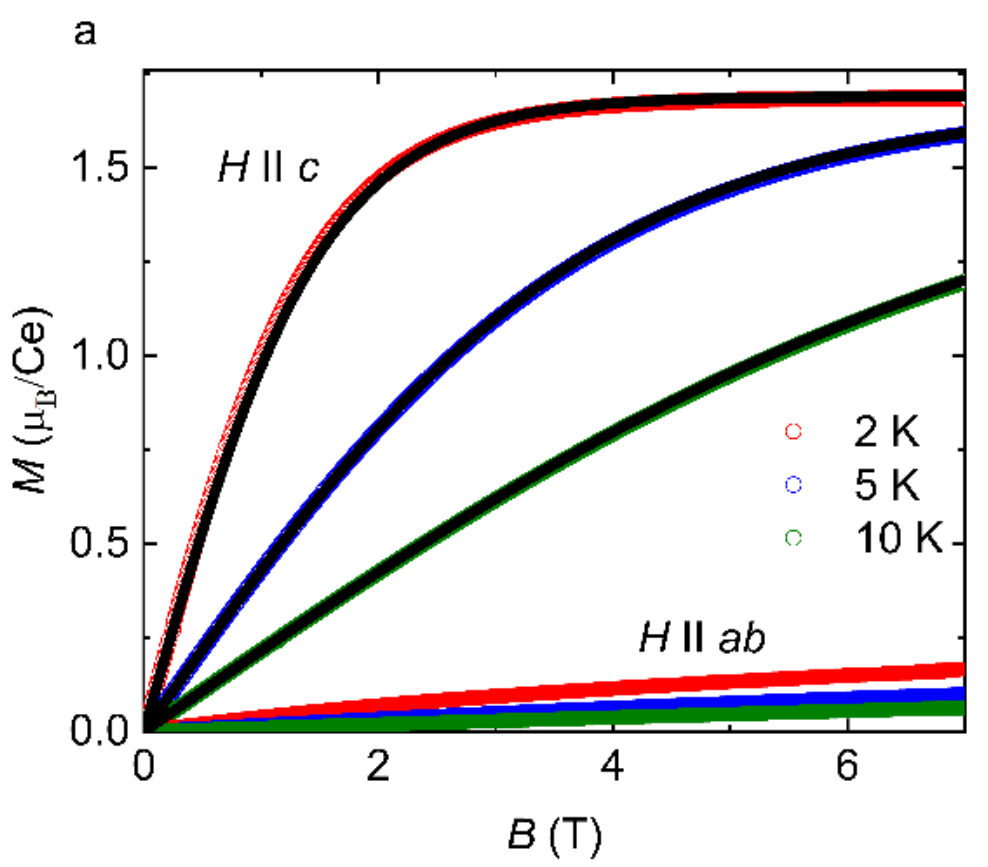

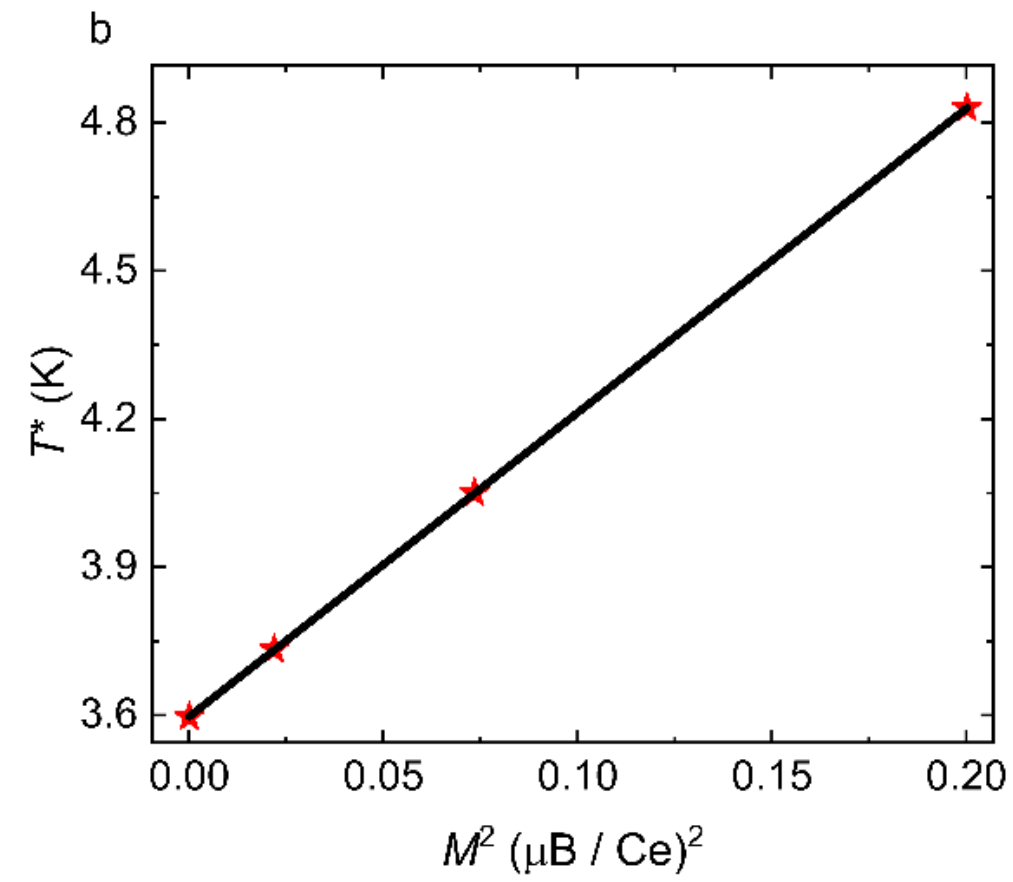

**Figure2.**

*(a) Isothermal magnetization curves M(H) measured at 2 K, 5 K, and 10 K (symbols). Solid black lines show Brillouin-function fits for an effective J = ½ moment ($M_{sat}$ = 1.69 $\mu_B$, g = 3.9). (b) Plot of the extracted anomaly temperatures T* versus the squared magnetization $M^2(H, T^*)$, computed from the Brillouin-function fits. The solid line represents the linear regression demonstrating the proportionality $T^* \propto M^2$, consistent with biquadratic magnetoelectric coupling.*

The field-dependent dielectric anomaly can be captured by a Landau-type free-energy expansion that includes the lowest symmetry-allowed ME invariant [35]:

$$F = F_0 + \tfrac{1}{2}\,\alpha_P(T)\,P^2 + \tfrac{1}{2}\,\alpha_M(T)\,M^2 + \gamma\,P^2M^2 + \ldots \qquad (6)$$

Here $\alpha_P(T)$ and $\alpha_M(T)$ denote the bare electric-dipole and magnetic stiffnesses, and γ is the biquadratic coupling constant, consistent with the centrosymmetric and time-reversal-invariant space group $P6_3/mmc$. The temperature dependence of $\alpha_P(T)$ is obtained from the Barrett fit [Fig. 1(a)] and reduces to $\alpha_P(T) = (T - T_0)/C$ in the high-T limit ($T \gg T_1$), with $T_0$ = 4.18 K and C = 263.3 K. Inclusion of the biquadratic term yields a field-renormalized stiffness $\alpha_P^{eff}(T, H) = \alpha_P(T) + 2\gamma C\,M^2(H, T)$.

Although no phase transition occurs because quantum fluctuations suppress long-range order down to 0.3 K, the position of the broad dielectric minimum—where dε'/dT = 0—can be phenomenologically described within this Landau framework. The anomaly temperature T*(H) is defined by

$$\alpha_P^{eff}(T^*, H) = [(T^*(H) - T_0)/C] - 2\gamma M^2(H, T^*(H)) \approx 0, \quad (7)$$

which leads to

$$T^*(H) = T_0 + 2\gamma C M^2(H, T^*(H)). \quad (8)$$

This expression reproduces the observed linear scaling of $T^*$ with $M^2$ and fits the experimental data without invoking a true critical point.

Experimentally, $\Delta T^*(H) = T^*(H) - T_0$ scales linearly with $M^2$, with slope $b = 6.2$ K $\mu B^{-2}$ [Fig. 1(d)], yielding the biquadratic ME coupling constant

$$\gamma = b / (2C) \approx 6.2 / (2 \times 263.3) = 1.2 \times 10^{-2} \text{ K } \mu B^{-2}. \quad (9)$$

The exclusive renormalization of $\alpha_P$ explains why the dielectric anomaly shifts in temperature without major amplitude changes. At $T^*$, the system remains paramagnetic—Brillouin fits confirm that a modest field of 1 T already polarizes $Ce^{3+}$ moments—yet the broad dielectric dip persists up to 5 T. Hence, the coupling does not stem from magnetic ordering or critical spin correlations but from virtual crystal-field (CEF) transitions of the Kramers doublet that renormalize the dielectric stiffness. Similar renormalizations have been reported in $DyFeO_3$, where a $P^2M^2$ term shifts dielectric anomalies [8]; in spin–phonon-coupled $EuTiO_3$ [32, 36]; and in incipient ferroelectrics such as $SrTiO_3$ and $KTaO_3$, where quantum fluctuations modify dielectric stiffness [28].

This biquadratic ME coupling in $CeMgAl_{11}O_{19}$ fundamentally differs from canonical multiferroics or type-II multiferroelectricity often found in triangular-lattice antiferromagnets [37]. In $Cr_2O_3$, a linear P·M term appears below $T_N \approx 310$ K due to broken inversion symmetry [1]. In $EuTiO_3$, the permittivity anomaly at $T_N \approx 5.3$ K arises from spin–phonon coupling via a soft $T_{1u}$ mode [32]. In $(Eu,Ba,Sr)TiO_3$, exhibiting multiferroic quantum criticality, biquadratic or gradient couplings originate from coherent phonon softening [38, 39]. None of these mechanisms applies to $CeMgAl_{11}O_{19}$: centrosymmetry forbids linear couplings, no long-range magnetic order occurs down to 30 mK, and no global phonon softening is observed. The field-tunable dielectric anomaly therefore reflects a distinct microscopic mechanism.

A comparison with $EuAl_{12}O_{19}$ is instructive. Both compounds host frustrated $AlO_5$ dipoles responsible for quantum-paraelectric behavior. However, $Eu^{2+}$ (L = 0) lacks orbital degrees of freedom, so no spin–orbit-driven ME coupling arises, and the permittivity remains unaffected by magnetic fields [20]. This contrast highlights the essential role of orbital-active rare-earth ions in enabling magnetoelectric interactions in hexaaluminates.

The key distinction in $CeMgAl_{11}O_{19}$ lies in its orbital-active Kramers $Ce^{3+}$ ion situated in a $D_3h$ crystal field, which splits the J = 5/2 manifold into three Kramers doublets [21]. The well-isolated ground doublet is separated from the first excited state by $\Delta_{CEF} \approx 14.1$ meV ($\approx 164$ K), far above the anomaly scale. Virtual coupling to local $AlO_5$ dipoles of moment p = q δ (with δ ≈ 0.2 Å [21] and q ≈ 3e) renormalizes the dielectric stiffness as

$$\Delta\alpha_P \propto (p^2 / \Delta_{CEF})\, M \qquad 10)$$

which, when matched to the Landau form, gives

$$\gamma_{th} \approx (3e\,\delta)^2 / [2\, \Delta_{CEF}\, k_B\, \mu B^2] \approx 1.0 \times 10^{-2}\ K\, \mu B^{-2}, \qquad (11)$$

in excellent agreement with the experimental $\gamma_{exp}$. A magnetic field introduces Zeeman splitting ($\Delta_Z$) that slightly modifies the denominator to $\Delta_{CEF} \pm \Delta_Z$, producing only sub-percent corrections and preserving the quadratic $T^* \propto M^2$ scaling. In $EuAl_{12}O_{19}$, by contrast, the L = 0 $Eu^{2+}$ ion suppresses this matrix element, giving γ ≈ 0 and eliminating the field-tunable anomaly [20].

These results establish $CeMgAl_{11}O_{19}$ as a prototype for a new class of magnetoelectric materials: centrosymmetric quantum paraelectrics in which the coupling is fluctuation-driven, biquadratic in form, and mediated by spin–orbit-entangled Kramers doublets interacting with an antipolar dipole background. This mechanism differs from conventional spin–phonon or exchange-striction pathways [34] and underscores the pivotal role of orbital-active rare-earth ions. The quantitative match between $\gamma_{th}$ and $\gamma_{exp}$ offers a guiding principle for realizing similar coupling in other frustrated paraelectrics. Candidate systems include triangular-lattice compounds hosting orbital-active Kramers ions such as $Rb_3Yb(PO_4)_2$ [40] and $YbBO_3$ [41], where a comparable fluctuation-driven mechanism may operate.

Importantly, the ground state of $CeMgAl_{11}O_{19}$ combines quantum paraelectricity with a proximate quantum-spin-liquid regime, giving rise to a quantum magnetoelectric state in which magnetic and electric dipoles may become entangled. Characterizing this state will require simultaneous studies of low-energy spin and dielectric dynamics under crossed electric and magnetic fields, for which the recently established crystal-field scheme [21] provides a solid microscopic foundation.

## V. Conclusions

$CeMgAl_{11}O_{19}$ stands out as a rare example of a quantum-paraelectric triangular magnet in which magnetoelectric coupling emerges from dynamic fluctuations of a Kramers doublet. The dielectric permittivity follows the Barrett law down to approximately 25 K and develops a broad, field-tunable minimum near 3 K, while magnetization measurements reveal anisotropic, fluctuation-dominated moments. The strict linear scaling $T^* \propto M^2$, confirmed by Landau analysis, reflects a centrosymmetric $P^2M^2$ coupling. Together, these results establish $CeMgAl_{11}O_{19}$ as a prototype system for magnetoelectricity mediated by spin–dipole coupling through crystal-field effects.

Future experiments could further elucidate this coupling. Temperature-dependent x-ray magnetic circular dichroism (XMCD) down to 0.3 K or high-field studies (up to 9 T) could test the robustness of the $T^* \propto M^2$ scaling and probe Zeeman-related corrections. It would also be of interest to measure ε'(T) under an external electric field E, where a similar biquadratic magnetoelectric coupling proportional to $E^2$ is expected. Such studies would refine the present understanding of the mechanism and extend this framework to other frustrated quantum magnets.

**Acknowledgments**

We acknowledge financial support from Charles University in Prague through the PRIMUS research program (Grant No. PRIMUS/22/SCI/016), the Grant Agency of Charles University (Grant No. 438425), and the Czech Science Foundation (Project No. 24/10791S). This work was also supported by the Ministry of Education, Youth and Sports of the Czech Republic through the INTER-EXCELLENCE II INTER-ACTION program (Project No. LUABA24056). Crystal growth, structural analysis, and magnetic-property measurements were carried out at the Materials Growth and Measurement Laboratory (MGML, http://mgml.eu/), supported within the Czech Research Infrastructures program (Project No. LM2023065).

**Author Contributions**

S. Kumar carried out crystal growth and sample preparation, participated in all experiments, performed data analysis and interpretation, and wrote the manuscript.
G. Bastien initiated the project, contributed to magnetization, specific heat, and dielectric measurements, and supervised the research.
P. Proschek and M. Savinov performed dielectric measurements.
A. Eliáš contributed to crystal growth and magnetization measurements.
K. Załęski reproduced magnetometry on additional crystals.
M. Śliwińska-Bartkowiak organized and supervised all work performed in Poznań.
R. H. Colman supervised crystal growth and interpretation of magnetization data.
S. Kamba supervised dielectric experiments, contributed to data interpretation, and oversaw manuscript preparation.